\documentclass[twocolumn,preprintnumbers,amsmath,amssymb,amsfonts]{revtex4}

\usepackage{graphicx}
\usepackage{mathrsfs}
\usepackage{amssymb}
\usepackage{color}
\usepackage{ulem}
\usepackage{mathtools}
\usepackage{float}
\usepackage{multirow}
\usepackage{diagbox}

\begin{document}
	\title{Probing quantumness of superpositions of Gaussian states via Tsirelson probability}
\author{Yue Zhang}
\email{zhangyue115@amss.ac.cn} 
	\affiliation{State Key Laboratory of Mathematical Science, Academy of Mathematics and
		Systems Science, Chinese Academy of Sciences,  Beijing 100190, China\\
		School of Mathematical Sciences, University of Chinese
		Academy of Sciences, Beijing 100049, China}
\begin{abstract}
Quantifying nonclassicality in continuous-variable systems remains a fundamental problem in quantum information science. The Tsirelson probability, central to the Tsirelson precession protocol, is defined as the average probability that a precessing quadrature yields a positive outcome when measured at $K$ equally spaced times, with the classical bound given by $1/2 \pm 1/(2K).$ In this work, we adopt this probability as a symmetry-sensitive probe to assess the quantumness of superpositions of Gaussian states in harmonic oscillators. We derive analytical constraints on Tsirelson probability arising from rotational and parity symmetries, and show that parity symmetry establishes a universal relation between the maximal and minimal Tsirelson probabilities within parity-related state families. Furthermore, we prove that even-fold rotational circular states cannot exhibit Tsirelson violation due to their definite parity. These results reveal how geometric symmetries of Gaussian-state superpositions determine their nonclassical behavior and provide a symmetry-based framework for probing quantumness in
continuous-variable systems.
\vskip 0.3cm
\end{abstract}

\maketitle
	
\section{Introduction} 
Identifying and quantifying nonclassicality in infinite-dimensional continuous-variable systems remains a fundamental challenge in quantum information science, with important implications for both fundamental studies and emerging quantum technologies. In quantum optics, a bosonic state is conventionally regarded as nonclassical if it cannot be represented as a statistical mixture of coherent states~\cite{Wall1994,Scul1997}. This notion is formally characterized by the Glauber--Sudarshan $P$-function: states possessing a positive $P$-representation are classical, whereas non-positive or highly singular $P$-functions indicate nonclassicality~\cite{Glau1963,Suda1963}. However, the direct characterization of the $P$-function is often experimentally and theoretically challenging, motivating the development of alternative quantifiers of nonclassicality, including sub-Poissonian photon statistics~\cite{Mand1979}, Wigner negativity~\cite{Kenf2004}, operator-ordering sensitivity~\cite{nc14}, Wigner--Yanase skew information~\cite{Luo2019}, and other quantifiers~\cite{Voge2000,Rich2002,
Ryl2017,Yadi2018,Tan2019,Luo2020,Zhan2022,Oliv2025}.

While these approaches provide powerful tools for characterizing nonclassicality, many of them rely on phase-space reconstruction or full state tomography. An alternative route is to identify quantum states through violations of classical constraints. In this context, the Tsirelson probability associated with rotating quadrature measurements provides an operational witness of nonclassicality in harmonic oscillators~\cite{Tsir1980,Tsir2006}. For a system undergoing uniform precession, one may ask whether the probability of obtaining positive outcomes for quadrature measurements at equally spaced times can be explained by a classical joint probability distribution. For three equally spaced measurement settings, classical theories constrain this probability to the interval $[1/3,2/3]$. Quantum states, however, can violate this bound, revealing nonclassical correlations using only the statistics of a single observable measured at different times~\cite{Zaw2022}. Subsequent studies have extended this framework to broader settings and applications~\cite{Chen2024,Plav2024,Zaw2025,Zaw20251,Garg2026}.

The harmonic oscillator provides a natural platform for exploring Tsirelson probabilities, since Gaussian states constitute a fundamental and experimentally accessible class of continuous-variable states. The pure Gaussian states, squeezed states, represent one of the most important classes of bosonic states, both theoretically and experimentally~\cite{Pere1986,Dodo2002,
Weed2012,Schn2017}. Meanwhile, the superposition principle provides a fundamental route to generating genuinely quantum states that defy classical phase-space descriptions~\cite{Schr1935,Buze1995,Silv2008}. Superpositions of Gaussian states, including coherent-state superpositions, squeezed-state superpositions, and their combinations, exhibit diverse nonclassical properties~\cite{Yuen1976,Sand1989,Jans1993,Chef1998,Zure2001,Horo2016,Luo2024,Kowa2021,Tang2026,Shuk2023,Tang2025}, such as decoherence~\cite{Zure2001}, entanglement~\cite{Horo2016}, Wigner negativity~\cite{Kowa2021,Tang2026}, and sub-Poissonian statistics~\cite{Shuk2023,Tang2025}. These states therefore provide ideal candidates for investigating the limitations of quantumness witnesses. Understanding how the structure of such superpositions determines their quantum behavior is a central problem in continuous-variable quantum information, and the Tsirelson precession protocol offers a particularly suitable framework for addressing it. 

Motivated by the rotational symmetry underlying the Tsirelson
precession protocol, we investigate circular states and generalized circular states as symmetry-adapted superpositions of Gaussian states. These states inherit the geometric structure of the measurement scenario and allow analytical investigation of their Tsirelson probabilities. In this work, we establish general symmetry constraints on Tsirelson probabilities for rotationally structured states. We show that parity symmetry imposes a universal relation between maximal and minimal Tsirelson probabilities and that even-fold rotational circular states cannot exhibit Tsirelson violation due to their definite parity. Furthermore, we analytically evaluate Tsirelson probabilities for generalized circular states, providing a systematic framework for probing the quantumness of Gaussian-state superpositions.

The remainder of this paper is organized as follows. In Sec. II, we revisit the fundamentals about circular states and generalized circular states. In Sec. III, we investigate Tsirelson probabilities for rotationally invariant states and derive symmetry constraints. In Sec. IV, we evaluate Tsirelson probabilities of circular states, as well as, generalized circular states. We make a conclusion in Section V. Soem detailed mathematical derivations of the main results are provided in the Appendix.
\section{Circular states and generalized circular states}

The Tsirelson precession protocol is characterized by a discrete rotational structure in phase space. This naturally motivates the construction of quantum states whose phase-space geometry inherits the same rotational symmetry. Circular states, introduced in Ref.~\cite{Chef1998,Luo2024}, and generalized circular states, introduced in Ref.~\cite{Tang2026}, are symmetry-adapted superpositions of Gaussian states that naturally fit this structure. In this section, we review their definitions and key properties, which will serve as the basis for our subsequent analysis of their Tsirelson probabilities.

In quantum optics, coherent states are among the most important states, and are typically viewed as the most `classical' pure states~\cite{Glau1963,Suda1963}. Let 
$|\alpha \rangle=D_\alpha|0\rangle,$ denote the coherent states, where 
\begin{align}\label{disp1}
D_{\alpha} = e^{\alpha a^\dag - \alpha^* a},\qquad \alpha\in\mathbb C
\end{align}
is the displacement operator satisfying 
\begin{align}\label{disp2}
D_\alpha D_\beta=e^{\frac12(\alpha\beta^*-\alpha^*\beta)}D_{\alpha+\beta},\qquad \alpha,\beta\in\mathbb C.
\end{align}
Equivalently, coherent states are the eigenstates of the annihilation operator $a$: $a|\alpha \rangle =\alpha |\alpha \rangle.$ The annihilation operator $a$ and creation operator $a^\dag$ satisfy the canonical commutation relation $[a,a^\dagger ]={\bf 1}.$

Circular states are obtained by coherently superposing coherent states distributed uniformly on a circle in phase space~\cite{Chef1998,Luo2024}. Let
\begin{align}
w_d=e^{2\pi i/d},
\end{align}
which will be adopted consistently in what follows. The circular state $|\alpha_{d,n}\rangle$ with rotational order $d\in\mathbb N_+$ and sector index $n\in \mathbb{Z}_d:=\{0,1,\cdots,d-1\}$ is
defined as 
\begin{equation}\label{cir}
|\alpha_{d,n}\rangle=\frac1{\sqrt{{\mathcal N}_{\alpha,d,n}}} \sum_{k=0}^{d-1} w_d^{-kn}|w_d^k\alpha\rangle,\qquad \alpha\in\mathbb{C}^*,
\end{equation}
where $\mathbb{C}^*$ is the set of nonzero complex numbers, and
\begin{align}\label{norm1}
{\mathcal N}_{\alpha,d,n}=d\sum_{k=0}^{d-1} w_d^{kn}  e^{(w_d^{-k}-1)|\alpha|^2}
\end{align}
is the normalization constant determined by the nonorthogonality of coherent states. When $d=2,$ the circular states reduce to the Schr\"odinger cat states, and for $d=4,$ to various compass states. 

Several key properties characterize the circular states~\cite{Luo2024}. Notably, the circular states belong to distinct eigenspaces (rotational sectors) of the discrete rotation operator,
\begin{align}\label{prop0}
R_{-\frac{2\pi}d}  |\alpha_{d,n}\rangle=w_d^n|\alpha_{d,n}\rangle,
\end{align}
where 
\begin{align}\label{rota}
R_t=e^{-it  a^\dag a},\qquad t\in\mathbb R
\end{align}
is the rotation operator satisfying $R_t^\dag a R_t=e^{-it}a.$ Thus, the index $n$ labels different rotational sectors of the Hilbert space. More generally, for any integer $m$ that divides $d,$ $R_{-{2\pi}/m}  |\alpha_{d,n}\rangle=w_d^{nd/m}|\alpha_{d,n}\rangle.$ 
Additionally, they are orthonormal, $\langle \alpha_{d,n}|\alpha_{d,n'}\rangle=\delta_{nn'}$ with $\delta_{nn'}=1$ if $n=n',$ and $0$ otherwise, and we note that the annihilation operator acts as a shift operator, $a|\alpha_{d,n}\rangle=\alpha\sqrt{{\mathcal N}_{\alpha,d,n-1}/{\mathcal N}_{\alpha,d,n}}|\alpha_{d,n-1}\rangle.$ Consequently, 
\begin{align}\label{prop1}
\langle \alpha_{d,n}|a|\alpha_{d,n}\rangle=0.
\end{align}

The squeezed coherent state can be defined as a displaced squeezed vacuum state ~\cite{Yuen1976}
\begin{equation} \label{eq:def_sq_coherent_state}
    |\alpha,\eta \rangle = D_{\alpha} S_{\eta} |0\rangle,\qquad \alpha,\eta\in\mathbb C,
\end{equation}
where the displacement operator $D_{\alpha}$ is given by Eq.~(\ref{disp1}) and $S_{\eta} = e^{\frac12 \eta a^{\dag 2} -\frac12 \eta^* a^2}$ is the squeezing operator with 
$$\eta:= r e^{i\theta},\qquad  r\geq0,\ \theta\in[0,2\pi).$$ 
For any two squeezed coherent states, their inner product in the Bargmann space is 
\begin{align}
\label{N12}
{\mathcal N}_{12}=\langle \alpha_1,\eta_1|\alpha_2,\eta_2\rangle
\end{align}
admits a closed analytical form, which is given in Appendix F. Rotate the squeezed coherent state by an angle of $-2\pi/d$ successively to create a set of $d$ states
\begin{equation}
    |\alpha_k,\eta_k\rangle: = R_{-\frac{2\pi}d} ^k |\alpha,\eta\rangle=|\alpha w_d^k, \eta w_d^{2k} \rangle, \qquad k\in \mathbb{Z}_d,
\end{equation}
whose displacement centers are symmetrically distributed on a circle in phase space.

Although coherent-state superpositions already exhibit rich nonclassical properties, extending the construction to general Gaussian states provides additional degrees of freedom through
squeezing. Therefore, by replacing the coherent-state constituents with rotated displaced squeezed states while preserving the underlying rotational structure, the generalized circular states were recently introduced in Ref.~\cite{Tang2026} as
\begin{equation}
    \label{gc}
    |{\boldsymbol g}_{d,n}\rangle = \frac{1}{\sqrt{\mathcal{N}_{{\boldsymbol g}, d, n}}} \sum_{k=0}^{d-1} w_d^{-kn} |\alpha_k, \eta_k\rangle,\qquad n\in \mathbb{Z}_d,
\end{equation}
where ${\boldsymbol g}=(\alpha,r,\theta) \in \mathbb{R}^+ \times \mathbb{R}^+ \times [0,2\pi),$ and the normalization constant is
\begin{align}\label{Ngdn}
   \mathcal{N}_{{\boldsymbol g}, d, n}= \sum_{k,l=0}^{d-1} w_d^{(l-k)n} N_{lk}
\end{align}
determined by the corresponding Gram
matrix
$$N_{lk}:=
\langle\alpha_l,\eta_l|\alpha_k,\eta_k\rangle=
\langle\alpha w_d^l,\eta w_d^{2l}|\alpha w_d^k,\eta w_d^{2k}\rangle,$$
whose analytical expression is given in Appendix F.

The generalized circular states preserve the discrete rotational structure of circular states as
\begin{equation}
    R_{-\frac{2\pi}m} |{\boldsymbol g}_{d,n}\rangle = w_d^{nd/m} |{\boldsymbol g}_{d,n}\rangle,
\end{equation}
for any integer $m$ that divides $d.$ Moreover  for any fixed $d$ and ${\boldsymbol g},$ the generalized circular states with various sector indexes are orthonormal, $
    \langle {\boldsymbol g}_{d,m} | {\boldsymbol g}_{d,n} \rangle = \delta_{mn}.$
Note that for any $d>1,$ \begin{align}\label{mean}
  &\  \langle {\boldsymbol g}_{d,n} | a | {\boldsymbol g}_{d,n} \rangle=\langle {\boldsymbol g}_{d,n} | a^\dag | {\boldsymbol g}_{d,n} \rangle= 0.
\end{align}

Generalized circular states enlarge the  variational space from coherent states to Gaussian-state superpositions. This additional flexibility allows us to investigate how squeezing and displacement jointly influence Tsirelson probabilities. In the following section, we first analyze the constraints imposed by rotational and parity symmetries on Tsirelson probabilities. These results provide general bounds independent of the specific Gaussian state construction introduced above.

\section{Symmetry constraints on Tsirelson probability}
The circular and generalized circular states reviewed above are constructed according to the discrete rotational symmetry of the Tsirelson protocol. Before analyzing these specific state families, we establish general constraints on Tsirelson probabilities arising from rotational and parity symmetries, which are independent of the particular Gaussian-state construction and apply to arbitrary quantum states satisfying the corresponding symmetry conditions.

In the Tsirelson protocol, a precessing observable is measured at a time chosen at random among three equally spaced times. Let $\textrm{Prob}_3$ denote the probability that the measurement outcome is positive. In classical physics, this probability obeys the Tsirelson inequality $1/3\leq \textrm{Prob}_3(\rho_{\rm cl})\leq 2/3$~\cite{Tsir2006}. But in the quantum case, Tsirelson’s inequality can be violated by some states due to their quantumness. For operators $Q(t)=(e^{-it}a +e^{it}a^\dag)\sqrt2$ (assuming $\hbar=1$), the relation 
$[Q(t),Q(t+\pi/2)]=i{\bf 1}$ holds. And note that
\begin{align}\label{rot}
Q(t)=R_t^\dag Q R_t=Q \cos t+P \sin t,
\end{align}
where $Q=Q(0)=(a+a^\dag)/\sqrt2$ and $P=Q(\pi/2)=(a-a^\dag)/(i\sqrt2)$ are the position and momentum operators. The score function is defined as $\Theta(x)=1$ for $x>0,$ $\Theta(x)=1/2$ for $x=0,$ and $\Theta(x)=0$ for $x<0$. The observable is measured randomly at a time from $\{2k\pi/3: \ k\in\mathbb Z\}$ and we check whether the outcome is positive or not, with the convention $\Theta(0)=1/2$. The Tsirelson probability in the quantum setting is
\begin{align*}
\textrm{Prob}_3(\rho)&=\frac13\Big\langle\Theta\big(Q(0)\big)+\Theta\big(Q(\frac{2\pi}{3} )\big)+\Theta\big(Q(\frac{4\pi}{3} )\big)\Big\rangle,
\end{align*}
where $\langle X\rangle={\rm tr}\rho X$ is the expectation value of any observable $X$ in the state $\rho.$ The bound of this probability over all quantum states is nontrivial, due to the noncommutativity of the three operators $Q(2k\pi/3),k\in\mathbb Z_3$. 

It was shown in Ref.~\cite{Zaw2022} that the maximum Tsirelson probability of the quantum harmonic oscillator satisfies $\textrm{Prob}_3(\rho)<0.822607$. Recent work has substantially tightened the bounds on the maximal Tsirelson probability to $0.709364\leq 
\sup_\rho \textrm{Prob}_3(\rho)\leq 0.730822,$ with numerical and asymptotic evidence suggesting
 $\sup_\rho\textrm{Prob}_3(\rho)\approx 0.709364$~\cite{Zaw2025}. However, the states that achieve the maximum Tsirelson probability $\textrm{Prob}_3$ are not known in a simple closed form. They are characterized by being invariant under a $2\pi/3$ rotation in phase space. Consequently, their Fock-space expansion only contains number states $|3k\rangle,\ k\in\mathbb N,$ i.e., those with photon numbers that are multiples of $3.$  These maximally violating states are obtained numerically by diagonalizing the Tsirelson operator within a truncated Fock space~\cite{Zaw2022,Zaw2025,Zaw20251}. 
 

The preceding characterization for $K=3$ can naturally generalize to arbitrary integer $K\geq3.$ Throughout this work, we fix the angular spacing between consecutive
measurements as
\begin{align}
\label{angle}
\varphi=-\frac{2\pi}{K},
\qquad K\geq3.
\end{align} 
For a quantum state $\rho$, the Tsirelson probability is defined as 
\begin{align}\label{Prob_d}
\textrm{Prob}_K(\rho)={\rm tr}\rho P_K,
\end{align}
where 
\begin{align}
P_K&=\frac 1K   \sum_{k=0}^{K-1} \Theta \big(Q(k\varphi )\big)
\end{align}
is the Tsirelson operator. Since $\textrm{Prob}_K(\rho)$ is linear in $\rho,$ and $0\leq\Theta(Q(t))\leq 1$ leading to $0\leq P_K\leq 1$ for any integer $K,$ 
$$0\leq\lambda_{\inf}(P_K)\leq\textrm{Prob}_K(\rho)\leq \lambda_{\sup}(P_K)\leq1,$$
where $\lambda_{\inf}(P_K),\lambda_{\sup}(P_K)$ are the spectral infimum and supremum of $P_K.$ Consequently, the extrema values of $\textrm{Prob}_K(\rho)$ over the convex set of quantum states are attained at pure states. Restricted in classical states, the corresponding Tsirelson inequality turns to be~\cite{Tsir2006} 
\begin{align}\label{cb}
\frac12-\frac1{2K}\leq \textrm{Prob}_K(\rho_{\rm cl})\leq \frac12+\frac1{2K},
\end{align}
where $K\geq 3$ is restricted to odd integers.

Recently, the Tsirelson operator has been analyzed from a group-theoretic perspective, revealing its cyclic rotational symmetry and invariant subspace structure \cite{Garg2026}. The following lemma reduces the optimization space via symmetry argument. 
\vskip0.2cm
\noindent\textbf{Lemma 1.} For any integer $K,$ the Tsirelson probability satisfies
\begin{align*}
\sup_{\rho}\textrm{Prob}_K(\rho)&= \sup_{\rho: R_\varphi \rho R_\varphi^\dag =\rho}\textrm{Prob}_K(\rho),\\
\inf_{\rho}\textrm{Prob}_K(\rho)&= \inf_{\rho: R_\varphi \rho R_\varphi^\dag =\rho}\textrm{Prob}_K(\rho). 
\end{align*}
\vskip0.2cm

This lemma indicates that the maximization or minimization of \(\textrm{Prob}_K(\rho)\) can be restricted, without loss of generality, to density operators invariant under the discrete rotation.

Note that $\Theta(x)=\big(1+{\rm sgn}(x)\big)/2,$ where ${\rm sgn}$ is the sign function. Then it holds that 
\begin{align}
\textrm{Prob}_K(\rho)=\frac12+\frac 1{2K} \sum_{k=0}^{K-1} \Big\langle  {\rm sgn}\big(Q(k\varphi )\big)\Big\rangle.
\end{align}
For $K=3$, probability $\textrm{Prob}_3(|\psi\rangle)$ for pure state $|\psi\rangle$ reduces to the original definition of the Tsirelson probability~\cite{Tsir2006}. 

The foregoing discussion raises a natural question: which states attain the extrema values of $\textrm{Prob}_K(\rho)$? While a complete characterization remains an open problem, the above lemmas provides a useful step toward understanding this question. To calculate the probability for a specific quantum state, the following lemmas and corollaries are useful.
\vskip0.2cm
\noindent\textbf{Lemma 2.} For any $K$-fold rotationally invariant state satisfying 
$$R_\varphi \rho R_\varphi^\dag =\rho,$$ 
the Tsirelson probability defined by Eq.~(\ref{Prob_d}), satisfies 
\begin{align}
\textrm{Prob}_K(\rho)&=\frac12+\frac12\big\langle {\rm sgn}(Q)\big\rangle.
\end{align}
\vskip0.2cm

This result shows that rotational invariance reduces the Tsirelson probability to a single quadrature expectation value. Directly, we have the following corollary.
\vskip0.2cm
\noindent\textbf{Corollary 1.} For any rotationally invariant state satisfying 
$R_\phi \rho R_\phi^\dag =\rho,$ where the measurement spacing satisfies $$\varphi=q\phi,\qquad q\in\mathbb N_+,\phi=-2\pi/d,d=qK,$$ the Tsirelson probability defined by Eq.~(\ref{Prob_d}) satisfies 
\begin{align*}
\textrm{Prob}_K(\rho)&=\frac12+\frac12\big\langle {\rm sgn}(Q)\big\rangle.
\end{align*}
\vskip0.2cm
\noindent\textbf{Lemma 3.} 
For a state family $\rho(\xi)$ satisfying the parity covariance relation 
$$\Pi\rho(\xi)\Pi^\dag =\rho(-\xi),$$
where 
\begin{align}\label{parity}
\Pi=e^{i\pi a^\dagger a}
\end{align}
is the parity operator, the Tsirelson probability obeys 
\begin{align*}
{\rm Prob}_{K}\big(\rho(\xi)\big)
=1-{\rm Prob}_{K}\big(\rho(-\xi)\big).
\end{align*}
Consequently if the range of \(\xi\) is invariant under \(\xi \to -\xi\),
\begin{align*}
\sup_{\xi}
{\rm Prob}_{K}\big(\rho(\xi)\big)
+
\inf_{\xi}
{\rm Prob}_{K}\big(\rho(\xi)\big)
=1 .
\end{align*}
\vskip0.2cm

This relation reflects the parity symmetry of Tsirelson probability, a fundamental symmetry between opposite phase-space directions, and independent of the detailed structure of the state family. Directly, we have the following corollary.
\vskip0.2cm
\noindent\textbf{Corollary 2.} 
\label{cor:even}
For a state family $\rho(\xi)$ satisfying the parity covariance relation $\Pi\rho(\xi)\Pi^\dag =\rho(-\xi)$ and
$$\rho(\xi)=\rho(-\xi),$$
the Tsirelson probability is always
\[
{\rm Prob}_K\big(\rho(\xi)\big)=\frac12,
\]
which lies at the center of the classical interval and therefore
cannot violate the Tsirelson bound.

The above results reveal that rotational and parity symmetries impose strong restrictions on Tsirelson probabilities. In particular, symmetry alone can prohibit violations for certain
even-fold circular-state families, whereas odd-fold rotational
symmetry does not impose the same restriction. These observations motivate the investigation of generalized circular states with additional Gaussian degrees of freedom in the following section.

\section{Tsirelson probability of generalized circular states}


The symmetry analysis in the previous section establishes general constraints on Tsirelson probabilities. We now consider circular states and generalized circular states as
superpositions of $d$ Gaussian pure states. In this section, we derive an analytical expression for the Tsirelson
probability of circular states and generalized circular states, and investigate how
displacement and squeezing parameters influence their quantumness.

\subsection{Circular states}

For the circular states $|\alpha_{d,n}\rangle$ defined by Eq.~(\ref{cir}), 
$$R_\varphi  |\alpha_{d,n}\rangle\langle \alpha_{d,n}|R_\varphi^\dag=|\alpha_{d,n}\rangle\langle \alpha_{d,n}|,\qquad d=qk,$$
direct from Eq.~(\ref{prop0}). Analytic calculation in Appendix D leads to 
\begin{align}\label{sgn}
\langle \alpha_{d,n}| {\rm sgn}(Q)|\alpha_{d,n}\rangle=\frac 1{{\mathcal N}_{\alpha,d,n}} \sum_{l,k=0}^{d-1} w_d^{(l-k)n} \mu_{lk},
\end{align}
with 
\begin{align}\label{mu}
\nonumber \mu_{lk}&:=\langle w_d^l\alpha|{\rm sgn}(Q)  |w_d^k\alpha\rangle\\
&=e^{(w_d^{k-l}-1)|\alpha|^2}h\big(\frac{w_d^{*l}\alpha^*+w_d^{k}\alpha}{\sqrt2}\big)
\end{align}
and the complex error function defined by 
\begin{align}\label{hz}
h(z)=\frac2{\sqrt\pi} \int_0^z e^{-t^2}{\rm d}t.
\end{align}
Comibined with lemma 2, we have the following proposition.

\noindent\textbf{Proposition 1.} For the circular states $|\alpha_{d,n}\rangle$ defined by Eq.~(\ref{cir}), $\alpha\in\mathbb C^*,d=qK,n\in\mathbb Z_d,$ and  $q,K\in\mathbb N_+,$ 
one has
$$\langle\alpha_{d,n}| \sum_{k=0}^{d-1} Q(\varphi k)|\alpha_{d,n}\rangle=0,\qquad d\geq2,$$
and the Tsirelson probability is 
\begin{align}
\nonumber \textrm{Prob}_K(|\alpha_{d,n}\rangle)=\frac12+\frac 1{2{\mathcal N}_{\alpha,d,n}} \sum_{l,k=0}^{d-1} w_d^{(l-k)n}  \mu_{lk},
\end{align}
where $\mu_{lk}$ is defined by Eq.~(\ref{mu}), $w_d=e^{2\pi i/d},$ and ${\mathcal N}_{\alpha,d,n}$ is defined by Eq.~(\ref{norm1}).
\vskip0.2cm

\begin{figure}[h]
    \centering \includegraphics[width=0.49\textwidth]{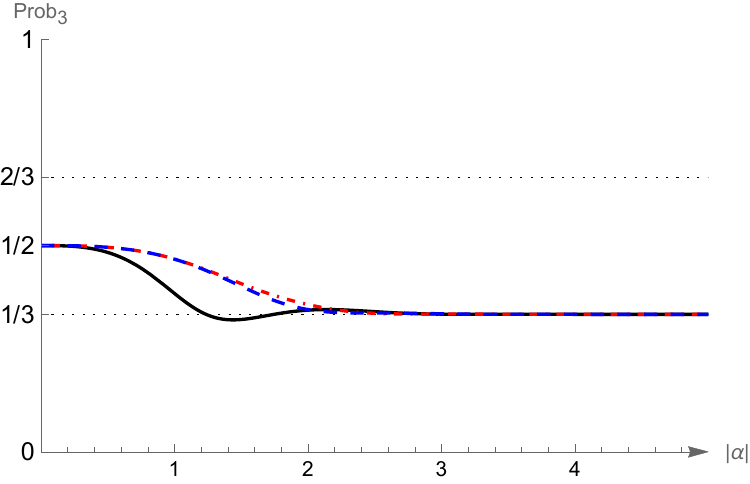}
    \caption{The Tsirelson probabilities for the circular states
$|\alpha_{3,0}\rangle$, $|\alpha_{3,1}\rangle$, and
$|\alpha_{3,2}\rangle$, shown as the black solid, red dash-dotted,
and blue dashed lines, respectively, with $\alpha$ restricted to the
positive real axis.}
    \label{fig0}
\end{figure}

\begin{figure}[h]
    \centering \includegraphics[width=0.49\textwidth]{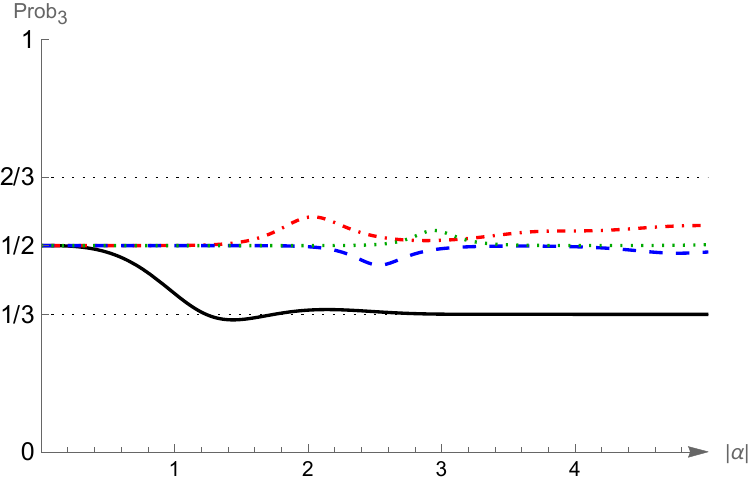}
\caption{Tsirelson probabilities for the circular states $|\alpha_{3,0}\rangle$, $|\alpha_{9,0}\rangle$, $|\alpha_{15,0}\rangle$, and $|\alpha_{21,0}\rangle$, shown as black solid, red dash-dotted, blue dashed, and green dotted lines, respectively, with $\alpha$ restricted to the positive real axis.}
    \label{fig02}
\end{figure}

\begin{figure}[h]
    \centering \includegraphics[width=0.49\textwidth]{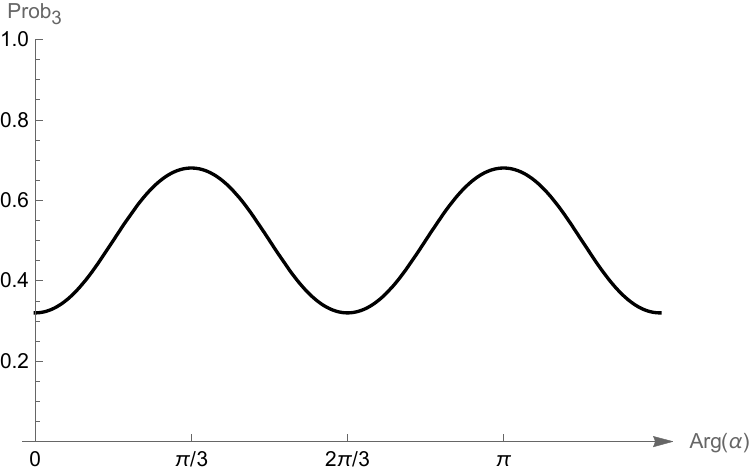}
    \caption{The Tsirelson probabilities for the circular states
$|\alpha_{3,0}\rangle$ as a function of the phase of $\alpha$,
with $|\alpha|\approx1.43892$.}
    \label{fig01}
\end{figure}

For $K=d=3$ and $n=0,1,2$, 
the Tsirelson probabilities of the circular states are plotted in Fig.~\ref{fig0}. Restricting
$\alpha$ to the positive real axis, the minimum value $\inf_{\alpha\in{\mathbb C}^*}\textrm{Prob}_3(|\alpha_{3,0}\rangle)\approx0.3199<1/3,$ is attained at $\alpha\approx1.43892.$ Fixing $K=3$ and $n=0,$ we consider circular states with $d=qK=3,9,15,21.$ As shown in Fig.~\ref{fig02}, the probability
violates the classical Tsirelson bound only for $d=3$ among the cases considered. 
\vskip0.2cm
\noindent\textbf{Proposition 2.} 
For the circular states $|\alpha_{d,n}\rangle$ defined by Eq.~(\ref{cir}), the parity transformation acts covariantly as
\begin{align*}
\Pi|\alpha_{d,n}\rangle
=|-\alpha_{d,n}\rangle ,
\end{align*}
where $\Pi$ is given by Eq.~(\ref{parity}). Consequently, the
Tsirelson probability satisfies
\begin{align*}
{\rm Prob}_{K}(|-\alpha_{d,n}\rangle)
=
1-
{\rm Prob}_{K}(|\alpha_{d,n}\rangle).
\end{align*}
Since the parameter domain $\alpha\in\mathbb C^*$ is invariant under
$\alpha\to-\alpha$, it follows that
\begin{align*}
\sup_{\alpha}
{\rm Prob}_{K}(|\alpha_{d,n}\rangle)
+
\inf_{\alpha}
{\rm Prob}_{K}(|\alpha_{d,n}\rangle)
=1 .
\end{align*}
In particular for any integer $K$ and even integer $d,$ 
\begin{align}
\nonumber \textrm{Prob}_K(|\alpha_{d,n}\rangle)=\frac12,\qquad \alpha\in\mathbb C^*,n\in\mathbb Z_d.
\end{align}
\vskip0.2cm
Specially for $K=d=3$ and $n=0$, the probability function ${\rm Porb}_3 (|\alpha_{3,0}\rangle)$ is periodic in the phase of
$\alpha$ with period $2\pi/3$ as shown in Fig. \ref{fig01}, reflecting the underlying three-fold rotational symmetry. The minimum Tsirelson probability is attained at zero phase, while the maximum  is
attained at $(1+2k)\pi/3,$ with $|\alpha|\approx1.43892.$ 

\begin{figure}[h]
    \centering \includegraphics[width=0.49\textwidth]{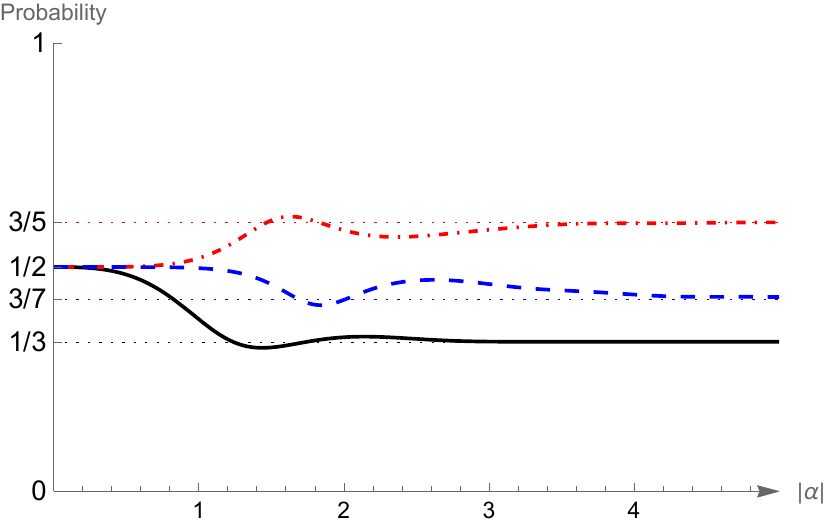}
    \caption{The Tsirelson probabilities for circular states $|\alpha_{3,0}\rangle,$   $|\alpha_{5,0}\rangle,$ and $|\alpha_{7,0}\rangle$ are plotted as the black solid, red dash-dotted, and blue dashed lines, respectively, with $\alpha$ restricted to the positive real axis.}
    \label{fig1}
\end{figure}

Correspondingly, the minimum and maximum are not independent optimization problems, as they are related by the parity symmetry established in Lemma~3. In the following, we restrict our analysis to the $n=0$ sector and real $\alpha\geq0$. For circular states with
$n=0$ and $d=K=3,5,7$, we find that the probability violates the classical Tsirelson bound. As shown in Fig.~\ref{fig1}, $\inf_{\alpha\geq0}\textrm{Prob}_3(|\alpha_{3,0}\rangle)\approx0.3199<1/3,$ $\sup_{\alpha\geq0}\textrm{Prob}_5(|\alpha_{5,0}\rangle)\approx0.6126>3/5,$ and $\inf_{\alpha\geq0}\textrm{Prob}_7(|\alpha_{7,0}\rangle)\approx0.4151<3/7.$ However, the violations achieved by these circular states remain
below the optimal quantum violation.

\subsection{Generalized circular states}

More generally, for generalized circular state, the superpositions of pure Gaussian states defined by Eq.~(\ref{gc}), the following proposition holds.
\vskip0.2cm
\noindent\textbf{Proposition 3.} For the generalized circular states $|\boldsymbol{g}_{d,n}\rangle$ defined by Eq.~(\ref{gc}) with $\boldsymbol{g}\in{\mathbb R}_+^2\times[0,2\pi),$ $d=qK$, $n\in{\mathbb Z}_d,$ and $q,K\in\mathbb N_+,$ one has
\begin{align}\label{mean2}
\langle {\boldsymbol g}_{d,n}| \sum_{k=0}^{d-1} Q(\varphi k) |{\boldsymbol g}_{d,n}\rangle =0,\qquad d\geq2.
\end{align}
Moreover, the Tsirelson probability is 
\begin{align*}
\textrm{Prob}_K(|{\boldsymbol g}_{d,n}\rangle)=\frac12+\frac 12\frac{\sum_{l,k=0}^{d-1}w_d^{(l-k)n} M_{lk}}{\sum_{l,k=0}^{d-1} w_d^{(l-k)n} N_{lk}},
\end{align*}
where $N_{lk}=\langle\alpha_l,\eta_l|
\alpha_k,\eta_k\rangle $ and $M_{lk}=\langle\alpha_l,\eta_l|{\rm sgn}(Q)|\alpha_k,\eta_k\rangle$ can be evaluated analytically using the Gaussian integral representation of the sign function and explicitly expressed in Appendix F.
\vskip0.2cm
Remarkably, as $r\to0,$ we have ${\rm th}r\to0,$ and thus  
$$N_{lk}\to\langle\alpha_l|\alpha_k\rangle,\qquad M_{lk}\to\mu_{lk}.$$ 
Consequently,
$$ \textrm{Prob}_K(|{\boldsymbol g}_{d,n}\rangle)\to\textrm{Prob}_K(|\alpha_{d,n}\rangle).$$ 
Thus, in the zero-squeezing limit, the generalized expression for the Tsirelson probability given in Proposition 3 reduces to the corresponding expression for the circular states given in Proposition 1.

\begin{figure}[h]
    \centering    \includegraphics[width=0.49\textwidth]{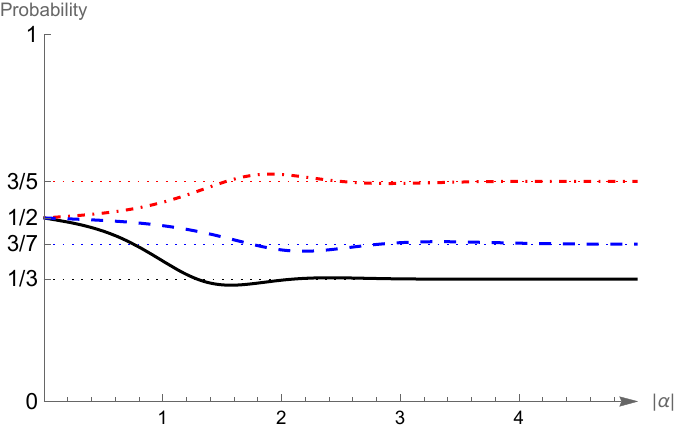}
    \caption{Tsirelson probabilities for the generalized circular states $|\boldsymbol{g}_{3,0}\rangle,$   $|\boldsymbol{g}_{5,0}\rangle,$ and $|\boldsymbol{g}_{7,0}\rangle,$ as functions of the displacement amplitude $|\alpha|$ for squeezing angle $\theta=0,$ are plotted as the black solid, red dash-dotted, and blue dashed lines, respectively. The squeezing strengths are optimized jointly with \(\alpha\), yielding \(r=0.1571\), \(0.3359\), and \(0.4617\), respectively.}
    \label{fig2}
\end{figure}
For the generalized circular states with $n=\theta=0$ and $d=K=3,5,7$, we numerically optimize the Tsirelson probability over the displacement amplitude $|\alpha|$ for different squeezing strengths $r$. We find violations of the classical Tsirelson bounds given by Eq.~(\ref{cb}). As shown in Fig. \ref{fig2}, for $d=K=3$, the infimum of the probability is obtained at $r=0.1571$, yielding
$\inf_{\alpha}\textrm{Prob}_3(|\boldsymbol{g}_{3,0}\rangle)\approx0.3167<1/3.$ For $d=5, r=0.3359$, the supremum exceeds the classical upper bound, $\sup_{\alpha}\textrm{Prob}_5(|\boldsymbol{g}_{5,0}\rangle)\approx0.6194>3/5.$ For $d=7, r=0.4617$, the infimum falls below the classical lower bound, $\inf_{\alpha}\textrm{Prob}_7(|\boldsymbol{g}_{7,0}\rangle)\approx0.4093<3/7.$ 

Compared with the corresponding circular states, the additional
squeezing degree of freedom enhances the magnitude of the Tsirelson violation for all three values of $d$ considered. This demonstrates that generalized circular states provide a richer parameter space for exploring nonclassicality, with the squeezing strength offering an additional degree of freedom that can further enhance the violation.

\section{Conclusion}
In this work, we have investigated the quantumness of superpositions of Gaussian states in harmonic oscillators through the Tsirelson probability. By exploiting the rotational structure of the Tsirelson precession protocol, we considered circular states and generalized circular states as symmetry-adapted families of Gaussian-state superpositions and analyzed their capability to exhibit nonclassical behavior.

For rotationally invariant states, we established general constraints on the Tsirelson probability arising from rotational and parity symmetries. In particular, we showed that parity symmetry leads to the relation in Lemma 3, which implies a universal relation between the maximal and minimal Tsirelson probabilities within parity-related state families. Furthermore, we demonstrated that circular states with even rotational order possess definite parity and therefore cannot exhibit Tsirelson violations. We further derived analytical expressions for the Tsirelson probabilities of generalized circular states, extending the analysis from coherent states to general Gaussian-state superpositions with displacement and squeezing degrees of freedom. These results show that the geometric structure and symmetry of Gaussian-state superpositions play a crucial role in determining their quantumness as quantified by Tsirelson probabilities.

Our work provides a symmetry-based framework for characterizing
nonclassicality in continuous-variable quantum systems. Beyond the specific state families considered here, the approach developed in this work may offer a useful perspective for identifying and understanding highly nonclassical states in infinite-dimensional systems.
\vskip 0.3cm
\noindent \textbf{Acknowledgements}. The author is grateful to Professor Shunlong Luo for discussions. This work was supported by the National Natural Science Foundation of China (Grant No. 12401609), the Youth Promotion Association of CAS (Grant No. 2023004).
\vskip 0.3cm
\noindent{\bf Data Availability Statement}. The data that support the findings of this study are available from the corresponding author upon reasonable request
\vskip 0.3cm
\noindent{\bf Conflict of interest}. {The authors declare that they have no conflict of interest.}
\vskip 0.3cm
\appendix
\noindent {\bf Appendix}	
\vskip 0.2cm
\noindent{\bf A. Proof of Lemma 1}
	\vskip 0.2cm  

Note that $R_\varphi^\dag P_K R_\varphi= \sum_{k=0}^{K-1} \Theta \big(Q((k+1)\varphi )\big)/d=P_K,$ i.e. the Tsirelson operator commute with rotation operator
$$[P_K,R_\varphi]=0.$$ 
Since \(P_K\) commutes with the \(K\)-fold rotation operator \(R_\varphi\), the Hilbert space decomposes into invariant rotational sectors. In specific, for any quantum state $\rho,$ define 
$$\rho_K=\frac 1K  \sum_{k=0}^{K-1}  R_\varphi^k \rho R_\varphi^{\dag k},$$
and therefore, ${\rm tr}\rho_K P_K=\sum_{k=0}^{K-1}  {\rm tr}R_\varphi^k \rho R_\varphi^{\dag k} P_K/K= {\rm tr} \rho  P_K,$
i.e.
$$\textrm{Prob}_K(\rho_K)=\textrm{Prob}_K(\rho),$$
which leads to lemma 1.

\vskip 0.2cm
\noindent{\bf B. Proof of Lemma 2}
\vskip 0.2cm  
First for any Hermitian operator $X,$ it holds that 
\begin{align}\label{Fourier}
{\rm sgn} (X)=-\frac i\pi \int_{\mathbb R} e^{itX}\frac{{\rm d}t}t. 
\end{align}
To prove this statement, note that the Fourier transformation of the function ${\rm sgn}(\lambda)$ is 
$$\widehat{\rm sgn} (t)=\int_{\mathbb R} e^{-i\lambda t}{\rm d}t=\frac2{it}, $$
and the inverse Fourier transformation is
$${\rm sgn} (\lambda)=-\frac i{\pi}{\rm PV}\int_{-\infty}^\infty e^{i\lambda t}\frac{{\rm d}t}t,$$
where the integral is understood in the Cauchy principal-value sense. For any Hermitian operator $X$ with spectral decomposition
$X=\int_{\mathbb R} \lambda {\rm d} E(\lambda),$ it holds that
\begin{align*}
{\rm sgn} (X)&=\int_{\mathbb R} {\rm sgn} (\lambda) {\rm d} E(\lambda)\\
&=-\frac i{\pi} \int_{\mathbb R} \int_{\mathbb R} e^{i\lambda t}\frac{{\rm d}t}t  {\rm d} E(\lambda) \\
&=-\frac i{\pi} \int_{\mathbb R} \int_{\mathbb R} e^{i\lambda t} {\rm d} E(\lambda)  \frac{{\rm d}t}t=-\frac i\pi \int_{\mathbb R} e^{itX}\frac{{\rm d}t}t, 
\end{align*}
which completes the proof of the statement (\ref{Fourier}).

Via the functional calculus, it holds that for any unitary operator $U,$   
\begin{align}
{\rm sgn}( U^\dag Q U)=-\frac i\pi \int_{\mathbb R} U^\dag e^{it   Q  } U\frac{{\rm d}t}t=U^\dag {\rm sgn}(Q)U.
\end{align}
Note that $Q(k\varphi )=R_{k\varphi }^\dag Q R_{k\varphi}=R_{kq\phi }^\dag Q R_{kq\phi}$ given by Eq.~(\ref{rot}), and $R_{k\varphi}=R_{\varphi}^k=R_{\phi}^{kq}$ from the group property of the rotation operators. Then for any $d$-fold rotationally invariant state satisfying $R_\phi \rho R_\phi^\dag =\rho,$ 
it holds that 
\begin{align*}
&\big\langle {\rm sgn} \big(Q(k\varphi )\big) \big\rangle
=\big\langle {\rm sgn}\big(R_\varphi^{\dag k} QR_\varphi^k\big)\big\rangle=\big\langle R_\varphi^{\dag k}  {\rm sgn}(Q)R_\varphi^k\big\rangle\\
&={\rm tr}\rho R_\varphi^{\dag k}  {\rm sgn}(Q)R_\varphi^k={\rm tr}R_\phi^{kq}\rho R_\phi^{\dag kq} {\rm sgn}(Q)=\big\langle {\rm sgn}(Q)\big\rangle.
\end{align*}
Therefore the probability is
\begin{align*}
\textrm{Prob}_K(\rho)&=\frac12+\frac1{2d}\Big\langle \sum_{k=0}^{d-1}{\rm sgn}\big(Q(\varphi k)\big)\Big\rangle\\ &=\frac12+\frac12\big\langle {\rm sgn}(Q)\big\rangle,
\end{align*}
which completes the proof of lemma 2 and corollary 1. 

\vskip 0.2cm
\noindent{\bf C. Proof of Lemma 3}
\vskip 0.2cm  

Since 
\begin{align*}
\Pi Q\Pi^\dagger=-Q ,
\end{align*} 
and the sign function preserves the spectral decomposition of the quadrature operator, we have
\begin{align*}
\Pi{\rm sgn}\big(Q(k\varphi)\big)\Pi^\dagger
=
-{\rm sgn}\big(Q(k\varphi)\big).
\end{align*}
For states that satisfy parity covariance
$$\Pi\rho(\xi)\Pi^\dag =\rho(-\xi),$$
we obtain
\begin{align*}
\begin{aligned}
{\rm tr} \rho(-\xi){\rm sgn}\big(Q(k\varphi)\big)
&={\rm tr} \rho(\xi)
\Pi^\dagger {\rm sgn}\big(Q(k\varphi)\big)\Pi
\\
&=-{\rm tr} \rho(\xi){\rm sgn}\big(Q(k\varphi)\big).
\end{aligned}
\end{align*}
Hence,
\begin{align*}
\begin{aligned}
{\rm Prob}_{K}\big(\rho(\xi)\big)
&=
\frac12
-\frac12
{\rm tr} \rho(-\xi){\rm sgn}\big(Q(k\varphi)\big)
\\
&=
1-{\rm Prob}_{K}\big(\rho(-\xi)\big).
\end{aligned}
\end{align*}
Consequently, 
\begin{align*}
\sup_{\xi}
{\rm Prob}_{K}\big(\rho(\xi)\big)
+
\inf_{\xi}
{\rm Prob}_{K}\big(\rho(\xi)\big)
=1 ,
\end{align*}
if the range of \(\xi\) is invariant under \(\xi \to -\xi\).

\vskip 0.2cm
\noindent{\bf D. Proof of Proposition 1}
\vskip 0.2cm 

For the circular states defined by Eq.~(\ref{cir}), it holds that 
$$\langle \alpha_{d,n}|Q(t)|\alpha_{d,n}\rangle=0$$
for any $d\geq2$ and $t\in[0,2\pi),$ which follows from the property of $|\alpha_{d,n}\rangle$ given by Eq.~(\ref{prop1}). Based on lemma 2, $R_\varphi  |\alpha_{d,n}\rangle=w_d^n|\alpha_{d,n}\rangle$ direct from Eq.~(\ref{prop0}), and the relation
\begin{align*}
&\quad \langle \alpha_{d,n}| {\rm sgn}(Q)|\alpha_{d,n}\rangle\\
&=\frac 1{{\mathcal N}_{\alpha,d,n}} \sum_{l,k=0}^{d-1} w_d^{(l-k)n} \langle w_d^l\alpha|{\rm sgn}(Q)  |w_d^k\alpha\rangle,
\end{align*}
the Tsirelson probability ${\rm Prob}_d(|\alpha_{d,n}\rangle)$ for any circular states is
\begin{align*}
&\quad {\rm Prob}_d(|\alpha_{d,n}\rangle)\\
&=\frac12+\frac12\langle \alpha_{d,n}| {\rm sgn}(Q)|\alpha_{d,n}\rangle\\
&=\frac12+\frac 1{2{\mathcal N}_{\alpha,d,n}} \sum_{l,k=0}^{d-1} w_d^{(l-k)n}  e^{(w_d^{k-l}-1)|\alpha|^2}h(\frac{w_d^{*l}\alpha^*+w_d^{k}\alpha}{\sqrt2}).
\end{align*}
Here the last equality holds because for any $\alpha,\beta\in\mathbb C,$ 
\begin{align*}
\langle\alpha | {\rm sgn}(Q)|\beta\rangle&= -\frac i{\pi}  \int_{\mathbb R}  \langle \alpha|e^{it Q}  |\beta\rangle \frac{{\rm d}t}t\\
&=-\frac i{\pi}e^{\alpha^*\beta -\frac12(|\alpha|^2+|\beta|^2)}  \int_{\mathbb R} e^{-\frac{t^2}4+\frac{it}{\sqrt2}(\alpha^*+\beta)} \frac{{\rm d}t}t\\
&=e^{\alpha^*\beta -\frac12(|\alpha|^2+|\beta|^2)} h(\frac{\alpha^*+\beta}{\sqrt2}),
\end{align*}
where $e^{it Q}=D_{it/\sqrt2},$ the second equality follows from the relation
$$\langle\alpha | D_{it/\sqrt2}|\beta\rangle=e^{\alpha^*\beta -\frac12(|\alpha|^2+|\beta|^2)-\frac{t^2}4+\frac{it}{\sqrt2}(\alpha^*+\beta)}$$ 
consequently from Eq.~(\ref{disp2}),
and the third equality involves the following important integral
\begin{align*}
I(z)&:= \int_{-\infty}^\infty e^{-\frac{t^2}4}\big(e^{izt} -1\big)\frac{{\rm d}t}t\\
&=I(0)+\int_0^z \frac{{\rm d}I(w)}{{\rm d}w} {\rm d}w\\
&=0+i\int_0^z \int_{-\infty}^\infty e^{-\frac14 t^2+iwt}{\rm d}t {\rm d}w  \\
&=i\int_0^z e^{-w_d^2}\int_{-\infty}^\infty e^{-\frac14 (t-2iw)^2}{\rm d}t {\rm d}w  \\
&=i 2\sqrt\pi \int_0^z e^{-w_d^2} {\rm d}w\\
&=i\pi h(z),
\end{align*}
with the complex error function defined by 
\begin{align*}
h(z)=\frac2{\sqrt\pi} \int_0^z e^{-t^2}{\rm d}t.
\end{align*}
Here the integrand of $I(z)$ is analytic at 
$t=0,$ because the numerator $(e^{izt} -1\big)$ vanishes linearly. Hence the integral converges for real $z$ and is extended to the complex plane by analytic continuation. Differentiation under the integral sign is justified by dominated convergence. $\int_{-\infty}^\infty e^{-(t-2iw)^2/4}{\rm d}t=2\sqrt\pi$ is evaluated by contour integration. Consequently, we have proved the  proposition 1.

\vskip 0.2cm
\noindent{\bf E. Proof of Proposition 2}
\vskip 0.2cm  
The circular states satisfy parity covariance,
\begin{align*}
\Pi|\alpha_{d,n}\rangle &=\frac1{\sqrt{{\mathcal N}_{\alpha,d,n}}} \sum_{k=0}^{d-1} w_d^{-kn}\Pi|w_d^k\alpha\rangle\\
&=\frac1{\sqrt{{\mathcal N}_{\alpha,d,n}}} \sum_{k=0}^{d-1} w_d^{-kn}|- w_d^k\alpha\rangle\\
&=|-\alpha_{d,n}\rangle ,
\end{align*}
and then Lemma 3 leads to
\begin{align*}
\begin{aligned}
{\rm Prob}_{K}(|-\alpha_{d,n}\rangle)
&=
\frac12
-\frac12
\langle\alpha_{d,n}|{\rm sgn}(Q)|\alpha_{d,n}\rangle
\\
&=
1-
{\rm Prob}_{K}(|\alpha_{d,n}\rangle).
\end{aligned}
\end{align*}
Consequently, 
\begin{align*}
\sup_{\alpha}
{\rm Prob}_{K}(|\alpha_{d,n}\rangle)
+
\inf_{\alpha}
{\rm Prob}_{K}(|\alpha_{d,n}\rangle)
=1 .
\end{align*}

For even $d$, one has
\begin{align*}
-1=e^{i\pi}=\omega^{d/2},
\end{align*}
and hence
\begin{align*}
-\alpha\omega^l
=
\alpha\omega^{l+d/2}.
\end{align*}
Changing the summation index according to
\begin{align*}
j=l+\frac d2 ,
\end{align*}
we obtain
\begin{align*}
\begin{aligned}
\Pi|\alpha_{d,n}\rangle
&=
\frac1{\sqrt{\mathcal N_{d,n}}}
\sum_{j=0}^{d-1}
\omega^{-n(j-d/2)}
|\alpha\omega^j\rangle
\\
&=
\omega^{nd/2}
|\alpha_{d,n}\rangle=(-1)^n|\alpha_{d,n}\rangle.
\end{aligned}
\end{align*}

Thus using the parity eigenvalue property,
\begin{align*}
\begin{aligned}
\langle\alpha_{d,n}|{\rm sgn}(Q)|\alpha_{d,n}\rangle
&=\langle\alpha_{d,n}|-\Pi^\dagger {\rm sgn}(Q)\Pi|\alpha_{d,n}\rangle
\\
&= -(-1)^{2n}\langle\alpha_{d,n}|
{\rm sgn}(Q)
|\alpha_{d,n}\rangle\\
&= -\langle\alpha_{d,n}|
{\rm sgn}(Q)
|\alpha_{d,n}\rangle.
\end{aligned}
\end{align*}
Therefore,
\begin{align*}
\langle {\rm sgn}(Q)\rangle=0 ,
\end{align*}
and by Lemma 2,
\begin{align*}
\operatorname{Prob}_{K}
(|\alpha_{d,n}\rangle)
=
\frac12 .
\end{align*}

\vskip 0.2cm
\noindent{\bf F. Proof of Proposition 3}
\vskip 0.2cm  
The inner product of any two squeezed coherent states is 
\begin{align}
\nonumber &{\mathcal N}_{12}=\langle \alpha_1,\eta_1|\alpha_2,\eta_2\rangle\\
\nonumber&= \frac{1} {\sqrt {u_{12}{\rm ch} r_1{\rm ch} r_2 }} e^{ \frac12(|\alpha_1|^2+|\alpha_2|^2) -\alpha_1\alpha_2^* }\times\\
& e^{\frac1{2u_{12}} \big( (\alpha_1-\alpha_2)^2 e^{-i\theta_1}{\rm th} r_1+ (\alpha_1^*-\alpha_2^*)^2 e^{i\theta_2}{\rm th} r_2-2| \alpha_1-\alpha_2|^2\big)},\label{N12d}
\end{align}
where 
\begin{align}
\label{u12} u_{12}=1-{\rm th} r_1{\rm th} r_2 e^{i(\theta_2-\theta_1)},
\end{align}
with  
$${\rm sh}x=\frac12(e^x-e^{-x}),\  {\rm ch}x=\frac12(e^x+e^{-x}),\   {\rm th}x=\frac{e^x-e^{-x}}{e^x+e^{-x}},$$
for any $x\in\mathbb R.$

Note that by Eq. (\ref{disp2}),
\begin{align*}
D_{it/\sqrt2} D_{\alpha_2}&=e^{\frac {it}{\sqrt2}{\rm Re}\alpha_2} D_{\alpha_2+it/\sqrt2}.
\end{align*}
For any $\alpha_1,\eta_1,\alpha_2,\eta_2\in\mathbb C$ and $t\in\mathbb R,$ we have 
\begin{align*}
\langle \alpha_1,\eta_1|e^{itQ}|\alpha_2,\eta_2\rangle&=\langle \alpha_1,\eta_1| D_{it/\sqrt2}|\alpha_2,\eta_2\rangle\\
&= e^{\frac{it}{\sqrt2}{\rm Re}(\alpha_2)}\langle \alpha_1,\eta_1|\alpha_2+\frac{it}{\sqrt2},\eta_2\rangle\\
&={\mathcal N}_{12}  e^{-\frac{\alpha_{12} t^2}4+\frac{i \beta_{12} t}{\sqrt2}}\\
&:=f(t),
\end{align*}
which is exactly a quadratic polynomial in $t$ inside the exponent, with the quadratic coefficient independent of $\alpha_1,\alpha_2$ and ${\rm Re}(\alpha_{12})>0$, 
where ${\mathcal N}_{12}$ and $u_{12}$ are given by Eq.~(\ref{N12d}) and Eq.~(\ref{u12}) respectively, and
\begin{align*}
\alpha_{12} &=\frac{(1+e^{i\theta_2}{\rm th} r_2)(1+e^{-i\theta_1}{\rm th} r_1)}{u_{12}},\\
\beta_{12} 
&=\alpha_1+\alpha_2^* +\frac{v_{12}}{u_{12}},\\
v_{12}&=(\alpha_1^*-\alpha_2^*)(1+e^{i\theta_2}{\rm th} r_2)+(\alpha_2-\alpha_1)(1+e^{-i\theta_1}{\rm th} r_1).
\end{align*}
Therefore
\begin{align*}
\langle \alpha_1,\eta_1|{\rm sgn} Q|\alpha_2,\eta_2\rangle =-\frac i\pi {\rm PV} \int_{-\infty}^\infty  f(t) \frac{{\rm d}t}{t}={\mathcal N}_{12} h\Big( \frac{\beta_{12}}{\sqrt{2\alpha_{12}}}\Big).
\end{align*}

Remarkably, \begin{align*}
f(0)&={\mathcal N}_{12}=\langle  \alpha_1,\eta_1|\alpha_2,\eta_2\rangle,\\
\frac{{\rm d} f}{{\rm d}t} (0)&=if(0)\beta_{12}/\sqrt2 =i\langle  \alpha_1,\eta_1|Q|\alpha_2,\eta_2\rangle,\\
\frac{{\rm d}^2 f}{{\rm d}t^2} (0)&=-f(0) (\beta_{12}^2+\alpha_{12})/2 =-\langle  \alpha_1,\eta_1|Q^2|\alpha_2,\eta_2\rangle,
\end{align*}
and thus 
\begin{align*}
\frac{\beta_{12}}{\sqrt{2\alpha_{12}}}=\frac{1}{\sqrt{2({\mathcal N}_{12} \langle  \alpha_1,\eta_1|Q^2|\alpha_2,\eta_2\rangle/\langle  \alpha_1,\eta_1|Q|\alpha_2,\eta_2\rangle^2 -1 ) } }.
\end{align*}

For convenience, we rewrite the generalized circular state as
\begin{align*}
|\boldsymbol g_{d,n}\rangle
=
\frac{1}{\sqrt{\mathcal N_{{\boldsymbol g}, d,n}}}
\sum_{l=0}^{d-1}
\omega^{-nl}
|\alpha_l,\eta_l\rangle.
\end{align*}
The normalization constant is $$\mathcal{N}_{{\boldsymbol g}, d, n}= \sum_{k,l=0}^{d-1} w_d^{(l-k)n} \langle\alpha_l,\eta_l|\alpha_k,\eta_k\rangle,$$ 
where
\begin{align*}
&\quad N_{lk}=\langle\alpha_l,\eta_l|\alpha_k,\eta_k\rangle=\langle\alpha w_d^l,\eta w_d^{2l}|\alpha w_d^k,\eta w_d^{2k}\rangle\\
&= \frac{1} {\sqrt {{\rm ch} r^2-{\rm sh} r^2 w_d^{2(k-l)} }}
e^{ |\alpha|^2(1-w_d^{l-k})}\times\\
& e^{\frac  {\alpha^2(w_d^l-w_d^k)^2 e^{-i\theta}w_d^{-2l}{\rm th} r+\alpha^{*2}(w_d^{-l}-w_d^{-k})^2 e^{i\theta}w_d^{2k}{\rm th} r-2|\alpha|^2|1-w_d^{k-l}|^2  } {2(1-w_d^{2(k-l)}{\rm th} ^2 r)}}.
\end{align*}
For any $d=qK$ with $K,q\in\mathbb N_+,n\in\mathbb Z_d,$ one has
$$R_\varphi |{\boldsymbol g}_{d,n}\rangle\langle{\boldsymbol g}_{d,n}| R_\varphi^\dag =|{\boldsymbol g}_{d,n}\rangle\langle{\boldsymbol g}_{d,n}|.$$ 
Based on lemma 2, the Tsirelson probability is 
\begin{align*}
&\quad \textrm{Prob}_K(|{\boldsymbol g}_{d,n}\rangle)\\
&=\frac12+\frac 1{2{\mathcal N}_{\boldsymbol{g},d,n}} \sum_{l,k=0}^{d-1} w_d^{(l-k)n} \langle\alpha_l,\eta_l|{\rm sgn}(Q)|\alpha_k,\eta_k\rangle,
\end{align*}
where 
\begin{align*}
M_{lk}&:=\langle\alpha_l,\eta_l|{\rm sgn}(Q)|\alpha_k,\eta_k\rangle\\
&=\langle\alpha w_d^l,\eta w_d^{2l}|{\rm sgn}(Q)|\alpha w_d^k,\eta w_d^{2k}\rangle=N_{lk} h\Big(\frac{b_{lk}}{\sqrt{2 a_{lk}}}\Big),
\end{align*}
with 
\begin{align*}
a_{lk}&=\frac{(1+w_d^{-2l}e^{-i\theta}{\rm th} r)(1+w_d^{2k}e^{i\theta}{\rm th} r)}{1-w_d^{2(k-l)}{\rm th} ^2 r},\\
b_{lk}&=\alpha w_d^l+\alpha^* w_d^{-k}+\frac{ \alpha^*(w_d^{-l}- w_d^{-k})(1+w_d^{2k}e^{i\theta}{\rm th} r)} {1-w_d^{2(k-l)}{\rm th} ^2 r}\\
&+\frac{\alpha(w_d^{k}- w_d^{l})(1+w_d^{-2l}e^{-i\theta}{\rm th} r) }{1-w_d^{2(k-l)}{\rm th} ^2 r},
\end{align*}
and the function $h(z)$ defined by Eq.~(\ref{hz}). 

Besides, Eq.~(\ref{mean2}) is a direct consequence of Eq.~(\ref{mean}), which implies
$$\langle {\boldsymbol g}_{d,n}| Q(t) |{\boldsymbol g}_{d,n}\rangle =0,\qquad t\in\mathbb R,$$ 
which actually completes the proof of Proposition 3.

\end{document}